\documentclass[aps,prb,amsmath,amssymb,superscriptaddress,reprint]{revtex4-1}
\usepackage{graphicx}

\begin{document}

\title{Adaptation of a commercial Raman spectrometer for multiline and broadband laser operation}

\author{G\'abor F\'abi\'an}
\affiliation{Department of Physics, Budapest University of Technology and Economics, Budafoki \'{u}t 8, H-1111 Budapest, Hungary}

\author{Christian Kramberger}
\affiliation{Faculty of Physics, University of Vienna, Strudlhofgasse 4, A-1090 Vienna, Austria}
\affiliation{Department of Mechanical Engineering, University of Tokyo, 7-3-1 Hongo, Bunkyo-ku, Tokyo 113-8656, Japan }

\author{Alexander Friedrich}
\affiliation{Faculty of Physics, University of Vienna, Strudlhofgasse 4, A-1090 Vienna, Austria}

\author{Ferenc Simon}
\affiliation{Department of Physics, Budapest University of Technology and Economics, Budafoki \'{u}t 8, H-1111 Budapest, Hungary}
\affiliation{Faculty of Physics, University of Vienna, Strudlhofgasse 4, A-1090 Vienna, Austria}

\author{Thomas Pichler}
\affiliation{Faculty of Physics, University of Vienna, Strudlhofgasse 4, A-1090 Vienna, Austria}







\begin{abstract}
A commercial single laser line Raman spectrometer is modified to accommodate multiline and tunable dye lasers, thus combining the high sensitivity of such single monochromator systems with broadband operation. Such instruments rely on high-throughput interference filters that perform both beam alignment and Rayleigh filtering. Our setup separates the dual task of the built-in monochromator into two independent elements: a beam
splitter and a long pass filter. Filter rotation shifts the transmission passband, effectively expanding the range of operation. Rotation of the filters has a negligible effect on the optical path, allowing broadband operation and stray light rejection down to 70-150 cm$^{-1}$. Operation is demonstrated on single-walled carbon nanotubes, for which the setup was optimized.

\end{abstract}

\maketitle
\section{Introduction}

Raman spectroscopy is a widespread and important tool in various fields of science from biology to physics. 
Commercial Raman spectrometers are usually equipped with a built-in laser and a setup optimized for this single laser line, 
resulting in stable operation but inherently narrow-band characteristics. 
The electronic, optical, and vibrational characterization of certain materials, such as single-wall carbon nanotubes 
(SWCNTs)\cite{DresselhausCNTRamanReview} requires measurements with a large number of laser lines \cite{KuzmanyEPJB} 
or with a tunable laser system \cite{FantiniPRL2004,TelgPRL2004}. 

Raman spectroscopy relies on the efficient
suppression of ``stray light'' photons with wavelengths close to that of the
exciting laser (e.g. from Rayleigh scattering) which dominate over the Raman signal by several orders of magnitude.
Operation down to Raman shifts of 100 cm$^{-1}$ is made possible in modern spectrometers with the use of 
interference Rayleigh filters (often referred to as notch filters). 
The transmission of these filters typically exceeds $80\,\%$ for the passband, this is significantly higher 
than for a classical subtractive double monochromator system. 
Although interference filters are manufactured for the most common laser lines only, rotation extends the range 
of filter operation. Thus the narrow-band constraint could be circumvented to allow broadband operation. However 
in most spectrometers, the interference filter has a dual role: it reflects the laser light to the sample and it
functions as a Rayleigh filter. Filter rotation changes the optical path of the excitation that can be corrected
for with tedious and time consuming readjustment only, effectively nullifying the advantage of the higher sensitivity.

In particular for the radial breathing mode of SWCNTs, the presence of the low energy ($\geq 100-150\,\text{cm}^{-1}$) \cite{RaoCNTRamanScience} Raman modes and the narrow
(FWHM $\sim 30 \,\text{meV}$) optical transition energies \cite{FantiniPRL2004} 
 pose several challenges to the instrumentation. A proper energy dependent Raman measurement requires a broadband spectrometer with
efficient stray light rejection. 

Herein, we describe the modification of a commercial Raman spectrometer with interference Rayleigh filters,
which enable broadband operation with relative ease. The improvement is based on replacing the built-in interference filter with a beam splitter and a separate interference filter. Thus the two functions of the filter are performed independently with no observable on influence the direction of the transmitted light.
The different behavior of the filter passband for the $S$ and $P$ \footnote{The S and P refer to polarizations which are perpendicular and parallel to the plane of incidence, respectively.}
polarizations under rotation is overcome by the application of polarization filters on the spectrometer input. The setup operates with polarizations which are
optimized when the so-called antenna effect of SWCNTs is taken into account, i.e. that the Raman light is polarized predominantly
along the polarization of the excitation \cite{SunAntennaEffect,JorioPhysRevLett85}.

\section{Spectrometer setup}

A high sensitivity, confocal single monochromator Raman system with a interference 
Rayleigh filter---such as described in the previous section---can be modified to enable 
broadband measurements with multiple laser lines or even with a tunable laser, which we 
demonstrate for a LabRAM commercial spectrometer (Horiba Jobin-Yvon Inc.) as an example. 
The key step in achieving the broadband operation was replacing the built-in interference 
filter, which acts as a beam splitter and a Rayleigh filter at the same time, with a 
combination of a simple beam splitter and a serarate interference filter. We note herein 
that this modification also enables a cost effective operation with usual laser wavelengths 
(such as e.g. the lines of an Ar/Kr laser) since no complicated filter realignments are required. We have to emphasize that the use of standard optical elements, which are non-specific to the spectrometer allows economic implementation for most spectrometer designs.

\begin{figure}[htp]
\begin{center}
\includegraphics[width=0.98\columnwidth]{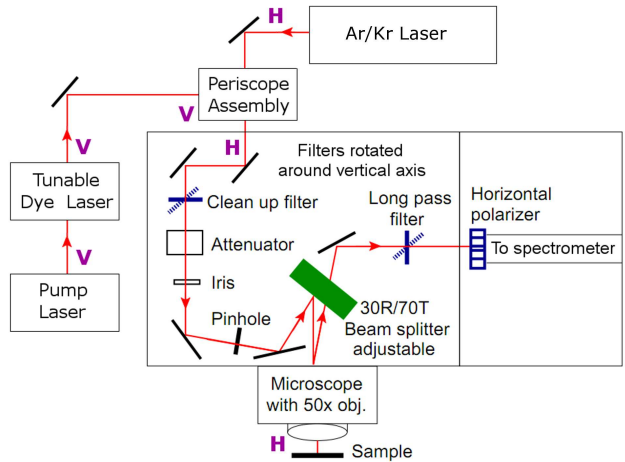}
\caption{Schematic diagram of the broadband configuration of the LabRAM spectrometer. V and H denote vertical and horizontal polarizations, respectively. The tunable source is a dye laser pumped with a 532 nm solid state laser. The laser light is aligned with the spectrometer using the periscope element, which also rotates the polarization to horizontal, if needed. The laser outputs are cleaned with a filter. The sample emits a nominally horizontally polarized light and the unwanted vertical polarization is filtered with the polarizer.}
\label{schem}
\end{center}
\end{figure}

The setup for the modified LabRAM spectrometer is shown in Fig. \ref{schem}. A multiline Ar/Kr laser (Coherent Inc., Innova C70C-Spectrum) and a dye laser (Coherent Inc., CR-590) pumped by a 532 nm 5 W solid state laser (Coherent Inc., Verdi G5) serve as excitation light sources. The former operates at multiple, well defined wavelengths while the latter allows fully tunable application.  In our case, the dye laser is operated in the 545-580 nm, 580-610 nm, and 610-660 nm wavelength ranges with three dyes: Rhodamin 110, Rhodamin 6G, and DCM Special, respectively.
The periscope allows beam alignment and sets the polarization of the excitation light to horizontal. 

In the case of the dye laser, the spurious fluorescent background of the laser output is filtered with short pass (``3rd Millennium filters'' for 580 and 610 nm from Omega Optical Inc.) and band pass (``RazorEdge'' for 568, 633, and 647 nm from Semrock Inc.) filters. For the clean-up of the multiline laser excitation band pass filters are used at the appropriate wavelengths (``RazorEdge'' for 458, 488, 515, 532, 568, 633, and 647 nm from Semrock Inc.) The light is directed toward the sample with a broadband beam splitter plate (Edmund Optics Inc., NT47-240) with 30 \% reflection and 70 \% transmission.  For both excitation sources a single, long pass interference edge filter (``RazorEdge'' for 458, 488, 515, 532, 568, 633, and 647 nm from Semrock Inc.) performs stray light rejection. The use of a short pass filter for laser clean-up and long pass filters for Rayleigh photon supression limits operation for the Stokes Raman range.

The long pass filter has double function in the original spectrometer: it mirrors the laser excitation to the sample and acts as a Rayleigh filter, quenching the stray light. In our construction, these two tasks are performed independently by a beam splitter and a long pass filter, respectively. The broadband beam splitter plate has 30 \% reflection and 70 \% transmission, thus only a small fraction of the Raman light is lost. The 70 \% excitation power loss on the beam splitter can be compensated by reducing the attenuation of the intensive laser beam, maintaining a constant irradiation density on the sample. The application of an anti-reflective coating to the back side of the plate prohibited the emergence of higher order reflections and standing waves (whose effect is known as ghosts) within the plate. The beam splitter plate is mounted on a finely adjustable 2-axis holder (Thorlabs Inc., VM1) with a home made mounting. The fine adjustment is required to set the light alignment properly with the spectrometer. Final fine adjustment is performed with the holder to maximize the Raman signal.

\begin{figure}[htp]
\begin{center}
\includegraphics[width=0.98\columnwidth]{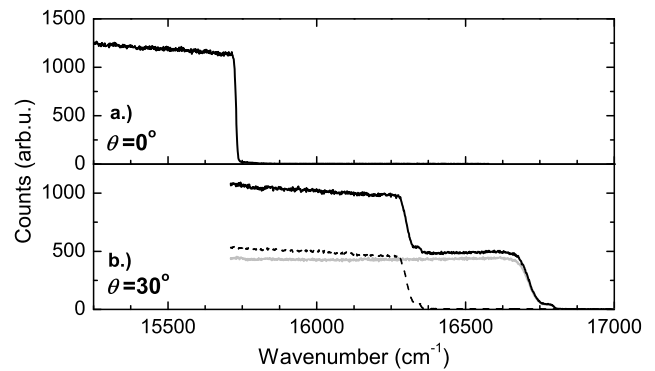}
\caption{Transmittance of the 633 nm long pass filter using unpolarized white light; a.) at normal incidence and b.) rotated by $30^{\circ}$. When polarization filters are used, the two parts of the double step feature (solid black line) are separated according to the $S$- and $P$-polarization (dashed black and solid gray lines, respectively). Note the broadening of filter transition width upon rotation.}
\label{LP}
\end{center}
\end{figure}

Increasing the incidence angle of the light changes the range of filter operation of the interference filters without the misalignment of the light. Thus filter rotation enables broadband operation. In Fig. \ref{LP}., we show the behavior of a 633 nm long pass filter at different incidence angles. The edge of transmission blue shifts upon rotation with respect to normal incidence. However, the shift is smaller for the $S$ than for the $P$ polarization; i.e. the shift is larger for the horizontally polarized light when the filter is rotated around a vertical axis. Vertical rotation of the long pass filter is more practical, meaning that the setup prefers horizontally polarized scattered (Raman) light as it is of the $P$ polarization, for which the edge shift is larger. For 1 inch apertures short and long pass filters rotation angles up to $30^{\circ}$ were used, yielding a blue shift of about 10 \%; the 0.5 inch aperture of band pass ``Razor edge'' filters limited the blue shift to about 5 \%. 
The width of the filter transition edge also broadens for larger incidence angles. This is defined as the maximum difference between the laser wavelength at which the attenuation exceeds optical density 6 and the filter edge-wavelength at the 50 \% transmission point. For the $30^{\circ}$ incidence, a fivefold increase in the transition width is observed when compared to the normal incidence, allowing operation down to $70-140\,\text{cm}^{-1}$.

For SWCNTs, the Raman light is polarized preferentially along the polarization of the excitation, this is due to a phenomenon called the antenna effect \cite{SunAntennaEffect,JorioPhysRevLett85}. We also verified that the LabRAM spectrometer itself is not polarization-sensitive in contrast to an older triple monochromator system. Therefore a horizontally polarized laser excitation is preferred which explains the polarizations used in our design. The less shifted $S$ (in our construction vertically) polarized stray light is removed with a polarization filter before the spectrometer input.

\section{Test measurements}

Test measurements of the broadband setup were carried out with the tunable dye laser on a HiPCO SWCNT sample (Carbon Nanotechnologies Inc., Houston, Texas), suspended in a 2 weight\% solution of SDBS (Sodium dodecyl benzene sulfonate) and water using sonication. 

We focused on the radial breathing mode (RBM) Raman range located below 400 $\text{cm}^{-1}$, which is commonly studied to characterize the diameter distribution in SWCNTs \cite{DresselhausTubes}. Carbon-tetrachloride was used for Raman shift correction and Raman intensity normalization such as in Ref. \cite{FantiniPRL2004}. The suspended HiPCO sample was placed in a glass cuvette under the objective of the built-in microscope (Olympus LMPlan 50x/0.50, inf./0/NN26.5, $\sim 1 \times 1 \,\mu \text{m}^2$ spot size) and the CCl$_4$ reference sample was placed into the macro cuvette holder. The LabRAM spectrometer allows to change between a macro and micro mode with a mirror resulting in stable and robust spectral shift calibration and intensity normalization as there is no need for further adjustments nor for sample exchange.

Laser excitation energies between 1.92 eV (648 nm) and 2.27 eV (545 nm) were covered with an energy resolution of about 12 meV ($\sim 100\,\text{cm}^{-1}$). The spectrometer was operated with a 600 grooves/mm grating and a liquid nitrogen cooled CCD with 1024 pixels along the spectral direction. This configuration yields a $\sim 1.3$ cm$^{-1}$ Raman shift resolution and $\sim 1800$ cm$^{-1}$ spectral range for 600 nm (both are wavelength dependent). Typical laser powers of 1-5 mW were used with no observable heating effects, due to the liquid nature of both samples.

\begin{figure}[htb]%
  \includegraphics[width=0.98\columnwidth]{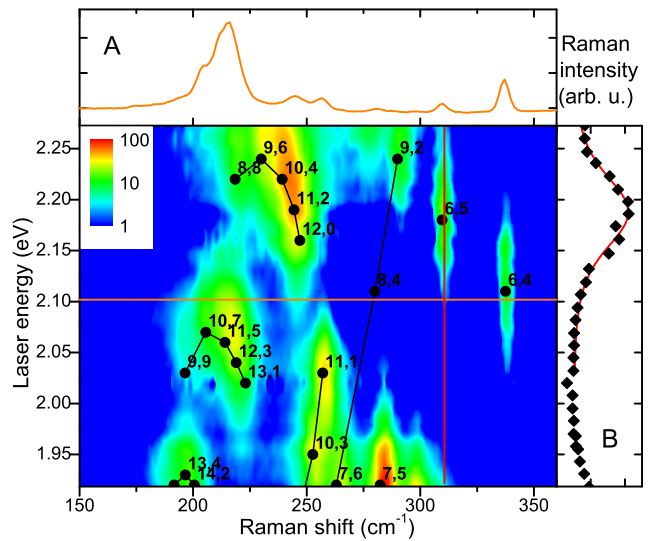}%
  \caption{Main plot: Raman map for the RBM range of a HiPCO/SDBS suspension measured with the broadband Raman setup.
Logarithmic scale shows the Raman intensity normalized to the maximum observed intensity.
Full circles denote data published in Ref. \cite{FantiniPRL2004}.
Inset A: A spectrum of the Raman map (horizontal line) at 592 nm (2.1 eV).
Inset B: Energy cross section of the Raman map (vertical line). The black diamonds correspond to the cross section at $310$ cm$^{-1}$ Raman shift, solid curve refers to a resonance Raman fit.
}
\label{map}
\end{figure}

An approximately 8 minute long measurement cycle consists of changing the dye laser wavelength, rotating of the laser clean-up and the long pass filters to the appropriate positions, shifting the spectrometer grating, nulling the spectrometer. Typical measurement times of 4 minutes for the sample and a few seconds for the reference yield an acceptable signal to noise ratio of about $300$. The sample and reference measurements followed each other immediately without moving the grating position, which led to an accurate Raman shift measurement. Additional time is needed for the filter exchange (a few minutes) and to change the laser dye and to readjust the beam alignment (about 1 hour). We note that once the dye laser is set and the light path is properly aligned with the spectrometer, no further realignment is required when the wavelength is changed, even though the filters are rotated and the optical path is only minutely modified.

In total, 9 hours were required to complete the energy dependent Raman experiments with the 29 laser lines including 2 laser dye exchanges. Measurements of a similar scale such as published previously using a triple monochromator spectrometer \cite{SimonPRB2006} last for about 2 weeks, mainly due to the approximately 50 times smaller S/N ratio and the need for a spectrometer realignment upon wavelength change.

Raman shift correction, intensity normalization, and the Raman map preparation was performed using a home-made software. Linear baseline correction was sufficient since no fluorescent response was encountered. Raman shift was corrected for with the carbon-tetrachloride Raman modes at 218 and 314 cm$^{-1}$. Intensity normalization using CCl$_4$ is required as it accounts for instrumental uncertainties such as a slight misalignment of the scattered light upon dye exchange. We checked the consistency of the normalization by measuring the same wavelength with different laser dyes after spectrometer realignment.

Fig. \ref{map}. shows the 2D contour plot of the Raman map, compiled from spectra such as shown in Fig. \ref{map}.A. The normalized Raman intensity is displayed on a logarithmic scale. A good agreement is observed between our data and the measurements in Ref. \cite{FantiniPRL2004}, whose resonance transition energies and Raman shifts are shown for the different SWCNT chiral indexes, $(n,m)$, with full circles. We do not observe the (8,4) SWCNT in our measurement, due the low intensity of the resonant Raman process species  \cite{Jorio2006APL}. Points corresponding to the same SWCNT families, i.e. when $2n+m$ is constant \cite{DresselhausTubes}, are connected by solid lines.

Vertical, i.e. energy cross section of the Raman map were obtained by averaging around a given Raman shift. Representative energy cross section data are shown in Fig. \ref{map}.B, 
along with fits using the resonance Raman theory \cite{FantiniPRL2004,SimonPRB2006}.
Fits yield transition energies and quasiparticle scattering rates in good agreement with typical literature values, especially considering that the different solvent environment slightly modifies the Raman transition energies \cite{FantiniPRL2004}.
We note that no further corrections were made to obtain the energy cross section data apart from the normalization by the reference. The result is therefore remarkably smooth in comparison with similar data published in Refs. \cite{FantiniPRL2004,TelgPRL2004,DoornPRB2008}. This is due to the robust and reproducible measurement of the reference sample and possibility of measuring Raman spectra at different wavelengths without spectrometer readjustment in between. The agreement shows the utility of the broadband arrangement with a clear advantage over previous results in terms of acquisition time.

\section{Conclusions}

In conclusion, we presented the broadband modification of a high sensitivity commercial Raman spectrometer. The improvement allows the use of both multiline and tunable dye lasers. The spectrometer performance is demonstrated on SWCNTs where such broadband measurements are inevitable to obtain meaningful insight into vibrational and electronic properties.

\begin{acknowledgments}
Work supported by the Austrian Science Funds (FWF) project Nr. P21333-N20, by the European Research Council Grant Nr. ERC-259374-Sylo, and by the New Hungary Development Plan Nr. T\'{A}MOP-4.2.1/B-09/1/KMR-2010-0002. CK acknowledges an APART fellowship (Nr. 11456)
of the Austrian Academy of Science. The authors acknowledge fruitful discussions with Dr. Neil Anderson from Semrock Inc. about the interference filters.
\end{acknowledgments}

merlin.mbs apsrev4-1.bst 2010-07-25 4.21a (PWD, AO, DPC) hacked

\begin{thebibliography}{12}%
\makeatletter
\providecommand \@ifxundefined [1]{%
 \@ifx{#1\undefined}
}%
\providecommand \@ifnum [1]{%
 \ifnum #1\expandafter \@firstoftwo
 \else \expandafter \@secondoftwo
 \fi
}%
\providecommand \@ifx [1]{%
 \ifx #1\expandafter \@firstoftwo
 \else \expandafter \@secondoftwo
 \fi
}%
\providecommand \natexlab [1]{#1}%
\providecommand \enquote  [1]{``#1''}%
\providecommand \bibnamefont  [1]{#1}%
\providecommand \bibfnamefont [1]{#1}%
\providecommand \citenamefont [1]{#1}%
\providecommand \href@noop [0]{\@secondoftwo}%
\providecommand \href [0]{\begingroup \@sanitize@url \@href}%
\providecommand \@href[1]{\@@startlink{#1}\@@href}%
\providecommand \@@href[1]{\endgroup#1\@@endlink}%
\providecommand \@sanitize@url [0]{\catcode `\\12\catcode `\$12\catcode
  `\&12\catcode `\#12\catcode `\^12\catcode `\_12\catcode `\%12\relax}%
\providecommand \@@startlink[1]{}%
\providecommand \@@endlink[0]{}%
\providecommand \url  [0]{\begingroup\@sanitize@url \@url }%
\providecommand \@url [1]{\endgroup\@href {#1}{\urlprefix }}%
\providecommand \urlprefix  [0]{URL }%
\providecommand \Eprint [0]{\href }%
\providecommand \doibase [0]{http://dx.doi.org/}%
\providecommand \selectlanguage [0]{\@gobble}%
\providecommand \bibinfo  [0]{\@secondoftwo}%
\providecommand \bibfield  [0]{\@secondoftwo}%
\providecommand \translation [1]{[#1]}%
\providecommand \BibitemOpen [0]{}%
\providecommand \bibitemStop [0]{}%
\providecommand \bibitemNoStop [0]{.\EOS\space}%
\providecommand \EOS [0]{\spacefactor3000\relax}%
\providecommand \BibitemShut  [1]{\csname bibitem#1\endcsname}%
\let\auto@bib@innerbib\@empty
\bibitem [{\citenamefont {Dresselhaus}\ \emph {et~al.}(2005)\citenamefont
  {Dresselhaus}, \citenamefont {Dresselhaus}, \citenamefont {Saito},\ and\
  \citenamefont {Jorio}}]{DresselhausCNTRamanReview}%
  \BibitemOpen
  \bibfield  {author} {\bibinfo {author} {\bibfnamefont {M.}~\bibnamefont
  {Dresselhaus}}, \bibinfo {author} {\bibfnamefont {G.}~\bibnamefont
  {Dresselhaus}}, \bibinfo {author} {\bibfnamefont {R.}~\bibnamefont {Saito}},
  \ and\ \bibinfo {author} {\bibfnamefont {A.}~\bibnamefont {Jorio}},\ }\href
  {\doibase DOI: 10.1016/j.physrep.2004.10.006} {\bibfield  {journal} {\bibinfo
   {journal} {Physics Reports}\ }\textbf {\bibinfo {volume} {409}},\ \bibinfo
  {pages} {47} (\bibinfo {year} {2005})}\BibitemShut {NoStop}%
\bibitem [{\citenamefont {Kuzmany}\ \emph {et~al.}(2001)\citenamefont
  {Kuzmany}, \citenamefont {Plank}, \citenamefont {Hulman}, \citenamefont
  {Kramberger}, \citenamefont {Gr\"uneis}, \citenamefont {Pichler},
  \citenamefont {Peterlik}, \citenamefont {Kataura},\ and\ \citenamefont
  {Achiba}}]{KuzmanyEPJB}%
  \BibitemOpen
  \bibfield  {author} {\bibinfo {author} {\bibfnamefont {H.}~\bibnamefont
  {Kuzmany}}, \bibinfo {author} {\bibfnamefont {W.}~\bibnamefont {Plank}},
  \bibinfo {author} {\bibfnamefont {M.}~\bibnamefont {Hulman}}, \bibinfo
  {author} {\bibfnamefont {C.}~\bibnamefont {Kramberger}}, \bibinfo {author}
  {\bibfnamefont {A.}~\bibnamefont {Gr\"uneis}}, \bibinfo {author}
  {\bibfnamefont {T.}~\bibnamefont {Pichler}}, \bibinfo {author} {\bibfnamefont
  {H.}~\bibnamefont {Peterlik}}, \bibinfo {author} {\bibfnamefont
  {H.}~\bibnamefont {Kataura}}, \ and\ \bibinfo {author} {\bibfnamefont
  {Y.}~\bibnamefont {Achiba}},\ }\href@noop {} {\bibfield  {journal} {\bibinfo
  {journal} {Eur. Phys. J. B}\ }\textbf {\bibinfo {volume} {22}},\ \bibinfo
  {pages} {307} (\bibinfo {year} {2001})}\BibitemShut {NoStop}%
\bibitem [{\citenamefont {Fantini}\ \emph {et~al.}(2004)\citenamefont
  {Fantini}, \citenamefont {Jorio}, \citenamefont {Souza}, \citenamefont
  {Strano}, \citenamefont {Dresselhaus},\ and\ \citenamefont
  {Pimenta}}]{FantiniPRL2004}%
  \BibitemOpen
  \bibfield  {author} {\bibinfo {author} {\bibfnamefont {C.}~\bibnamefont
  {Fantini}}, \bibinfo {author} {\bibfnamefont {A.}~\bibnamefont {Jorio}},
  \bibinfo {author} {\bibfnamefont {M.}~\bibnamefont {Souza}}, \bibinfo
  {author} {\bibfnamefont {M.~S.}\ \bibnamefont {Strano}}, \bibinfo {author}
  {\bibfnamefont {M.~S.}\ \bibnamefont {Dresselhaus}}, \ and\ \bibinfo {author}
  {\bibfnamefont {M.~A.}\ \bibnamefont {Pimenta}},\ }\href@noop {} {\bibfield
  {journal} {\bibinfo  {journal} {Phys. Rev. Lett.}\ }\textbf {\bibinfo
  {volume} {93}},\ \bibinfo {pages} {147406} (\bibinfo {year}
  {2004})}\BibitemShut {NoStop}%
\bibitem [{\citenamefont {Telg}\ \emph {et~al.}(2004)\citenamefont {Telg},
  \citenamefont {Maultzsch}, \citenamefont {Reich}, \citenamefont {Hennrich},\
  and\ \citenamefont {Thomsen}}]{TelgPRL2004}%
  \BibitemOpen
  \bibfield  {author} {\bibinfo {author} {\bibfnamefont {H.}~\bibnamefont
  {Telg}}, \bibinfo {author} {\bibfnamefont {J.}~\bibnamefont {Maultzsch}},
  \bibinfo {author} {\bibfnamefont {S.}~\bibnamefont {Reich}}, \bibinfo
  {author} {\bibfnamefont {F.}~\bibnamefont {Hennrich}}, \ and\ \bibinfo
  {author} {\bibfnamefont {C.}~\bibnamefont {Thomsen}},\ }\href@noop {}
  {\bibfield  {journal} {\bibinfo  {journal} {Phys. Rev. Lett.}\ }\textbf
  {\bibinfo {volume} {93}},\ \bibinfo {pages} {177401} (\bibinfo {year}
  {2004})}\BibitemShut {NoStop}%
\bibitem [{\citenamefont {Rao}\ \emph {et~al.}(1997)\citenamefont {Rao},
  \citenamefont {Richter}, \citenamefont {Bandow}, \citenamefont {Chase},
  \citenamefont {Eklund}, \citenamefont {Williams}, \citenamefont {Fang},
  \citenamefont {Subbaswamy}, \citenamefont {Menon}, \citenamefont {Thess},
  \citenamefont {Smalley}, \citenamefont {Dresselhaus},\ and\ \citenamefont
  {Dresselhaus}}]{RaoCNTRamanScience}%
  \BibitemOpen
  \bibfield  {author} {\bibinfo {author} {\bibfnamefont {A.}~\bibnamefont
  {Rao}}, \bibinfo {author} {\bibfnamefont {E.}~\bibnamefont {Richter}},
  \bibinfo {author} {\bibfnamefont {S.}~\bibnamefont {Bandow}}, \bibinfo
  {author} {\bibfnamefont {B.}~\bibnamefont {Chase}}, \bibinfo {author}
  {\bibfnamefont {P.}~\bibnamefont {Eklund}}, \bibinfo {author} {\bibfnamefont
  {K.}~\bibnamefont {Williams}}, \bibinfo {author} {\bibfnamefont
  {S.}~\bibnamefont {Fang}}, \bibinfo {author} {\bibfnamefont {K.}~\bibnamefont
  {Subbaswamy}}, \bibinfo {author} {\bibfnamefont {M.}~\bibnamefont {Menon}},
  \bibinfo {author} {\bibfnamefont {A.}~\bibnamefont {Thess}}, \bibinfo
  {author} {\bibfnamefont {R.}~\bibnamefont {Smalley}}, \bibinfo {author}
  {\bibfnamefont {G.}~\bibnamefont {Dresselhaus}}, \ and\ \bibinfo {author}
  {\bibfnamefont {M.}~\bibnamefont {Dresselhaus}},\ }\href@noop {} {\bibfield
  {journal} {\bibinfo  {journal} {Science}\ }\textbf {\bibinfo {volume}
  {275}},\ \bibinfo {pages} {187} (\bibinfo {year} {1997})}\BibitemShut
  {NoStop}%
\bibitem [{Note1()}]{Note1}%
  \BibitemOpen
  \bibinfo {note} {The S and P refer to polarizations which are perpendicular
  and parallel to the plane of incidence, respectively.}\BibitemShut {Stop}%
\bibitem [{\citenamefont {Sun}\ \emph {et~al.}(1999)\citenamefont {Sun},
  \citenamefont {Tang}, \citenamefont {Chen},\ and\ \citenamefont
  {Li}}]{SunAntennaEffect}%
  \BibitemOpen
  \bibfield  {author} {\bibinfo {author} {\bibfnamefont {H.~D.}\ \bibnamefont
  {Sun}}, \bibinfo {author} {\bibfnamefont {Z.~K.}\ \bibnamefont {Tang}},
  \bibinfo {author} {\bibfnamefont {J.}~\bibnamefont {Chen}}, \ and\ \bibinfo
  {author} {\bibfnamefont {G.}~\bibnamefont {Li}},\ }\href {\doibase DOI:
  10.1016/S0038-1098(98)00588-2} {\bibfield  {journal} {\bibinfo  {journal}
  {Solid State Communications}\ }\textbf {\bibinfo {volume} {109}},\ \bibinfo
  {pages} {365} (\bibinfo {year} {1999})}\BibitemShut {NoStop}%
\bibitem [{\citenamefont {Jorio}\ \emph {et~al.}(2000)\citenamefont {Jorio},
  \citenamefont {Dresselhaus}, \citenamefont {Dresselhaus}, \citenamefont
  {Souza}, \citenamefont {Dantas}, \citenamefont {Pimenta}, \citenamefont
  {Rao}, \citenamefont {Saito}, \citenamefont {Liu},\ and\ \citenamefont
  {Cheng}}]{JorioPhysRevLett85}%
  \BibitemOpen
  \bibfield  {author} {\bibinfo {author} {\bibfnamefont {A.}~\bibnamefont
  {Jorio}}, \bibinfo {author} {\bibfnamefont {G.}~\bibnamefont {Dresselhaus}},
  \bibinfo {author} {\bibfnamefont {M.~S.}\ \bibnamefont {Dresselhaus}},
  \bibinfo {author} {\bibfnamefont {M.}~\bibnamefont {Souza}}, \bibinfo
  {author} {\bibfnamefont {M.~S.~S.}\ \bibnamefont {Dantas}}, \bibinfo {author}
  {\bibfnamefont {M.~A.}\ \bibnamefont {Pimenta}}, \bibinfo {author}
  {\bibfnamefont {A.~M.}\ \bibnamefont {Rao}}, \bibinfo {author} {\bibfnamefont
  {R.}~\bibnamefont {Saito}}, \bibinfo {author} {\bibfnamefont
  {C.}~\bibnamefont {Liu}}, \ and\ \bibinfo {author} {\bibfnamefont {H.~M.}\
  \bibnamefont {Cheng}},\ }\href@noop {} {\bibfield  {journal} {\bibinfo
  {journal} {Phys. Rev. Lett.}\ }\textbf {\bibinfo {volume} {85}},\ \bibinfo
  {pages} {2617} (\bibinfo {year} {2000})}\BibitemShut {NoStop}%
\bibitem [{\citenamefont {Saito}\ \emph {et~al.}(1998)\citenamefont {Saito},
  \citenamefont {Dresselhaus},\ and\ \citenamefont
  {Dresselhaus}}]{DresselhausTubes}%
  \BibitemOpen
  \bibfield  {author} {\bibinfo {author} {\bibfnamefont {R.}~\bibnamefont
  {Saito}}, \bibinfo {author} {\bibfnamefont {G.}~\bibnamefont {Dresselhaus}},
  \ and\ \bibinfo {author} {\bibfnamefont {M.}~\bibnamefont {Dresselhaus}},\
  }\href@noop {} {\emph {\bibinfo {title} {Physical Properties of Carbon
  Nanotubes}}}\ (\bibinfo  {publisher} {Imperial College Press},\ \bibinfo
  {year} {1998})\BibitemShut {NoStop}%
\bibitem [{\citenamefont {Simon}\ \emph {et~al.}(2006)\citenamefont {Simon},
  \citenamefont {Pfeiffer},\ and\ \citenamefont {Kuzmany}}]{SimonPRB2006}%
  \BibitemOpen
  \bibfield  {author} {\bibinfo {author} {\bibfnamefont {F.}~\bibnamefont
  {Simon}}, \bibinfo {author} {\bibfnamefont {R.}~\bibnamefont {Pfeiffer}}, \
  and\ \bibinfo {author} {\bibfnamefont {H.}~\bibnamefont {Kuzmany}},\
  }\href@noop {} {\bibfield  {journal} {\bibinfo  {journal} {Phys. Rev. B}\
  }\textbf {\bibinfo {volume} {74}},\ \bibinfo {pages} {212411(R)} (\bibinfo
  {year} {2006})}\BibitemShut {NoStop}%
\bibitem [{\citenamefont {Jorio}\ \emph {et~al.}(2006)\citenamefont {Jorio},
  \citenamefont {Fantini}, \citenamefont {Pimenta}, \citenamefont {Heller},
  \citenamefont {Strano}, \citenamefont {Dresselhaus}, \citenamefont {Oyama},
  \citenamefont {Jiang},\ and\ \citenamefont {Saito}}]{Jorio2006APL}%
  \BibitemOpen
  \bibfield  {author} {\bibinfo {author} {\bibfnamefont {A.}~\bibnamefont
  {Jorio}}, \bibinfo {author} {\bibfnamefont {C.}~\bibnamefont {Fantini}},
  \bibinfo {author} {\bibfnamefont {M.~A.}\ \bibnamefont {Pimenta}}, \bibinfo
  {author} {\bibfnamefont {D.~A.}\ \bibnamefont {Heller}}, \bibinfo {author}
  {\bibfnamefont {M.~S.}\ \bibnamefont {Strano}}, \bibinfo {author}
  {\bibfnamefont {M.~S.}\ \bibnamefont {Dresselhaus}}, \bibinfo {author}
  {\bibfnamefont {Y.}~\bibnamefont {Oyama}}, \bibinfo {author} {\bibfnamefont
  {J.}~\bibnamefont {Jiang}}, \ and\ \bibinfo {author} {\bibfnamefont
  {R.}~\bibnamefont {Saito}},\ }\href {\doibase 10.1063/1.2162688} {\bibfield
  {journal} {\bibinfo  {journal} {Applied Physics Letters}\ }\textbf {\bibinfo
  {volume} {88}},\ \bibinfo {eid} {023109} (\bibinfo {year}
  {2006})}\BibitemShut {NoStop}%
\bibitem [{\citenamefont {Doorn}\ \emph {et~al.}(2008)\citenamefont {Doorn},
  \citenamefont {Araujo}, \citenamefont {Hata},\ and\ \citenamefont
  {Jorio}}]{DoornPRB2008}%
  \BibitemOpen
  \bibfield  {author} {\bibinfo {author} {\bibfnamefont {S.~K.}\ \bibnamefont
  {Doorn}}, \bibinfo {author} {\bibfnamefont {P.~T.}\ \bibnamefont {Araujo}},
  \bibinfo {author} {\bibfnamefont {K.}~\bibnamefont {Hata}}, \ and\ \bibinfo
  {author} {\bibfnamefont {A.}~\bibnamefont {Jorio}},\ }\href {\doibase
  10.1103/PhysRevB.78.165408} {\bibfield  {journal} {\bibinfo  {journal} {Phys.
  Rev. B}\ }\textbf {\bibinfo {volume} {78}},\ \bibinfo {pages} {165408}
  (\bibinfo {year} {2008})}\BibitemShut {NoStop}%
\end{thebibliography}
\end{document}